\documentclass{elsart}
\usepackage{epsfig}
    \newcommand{\ba}{\begin{eqnarray}}
    \newcommand{\ea}{\end{eqnarray}}
    \newcommand{\be}{\begin{equation}}
    \newcommand{\ee}{\end{equation}}

    \newcommand{\AmS}{{\protect\the\textfont2%
  A\kern-.1667em\lower.5ex\hbox{M}\kern-.125emS}}

\newcommand{\bx}{{\bf x}}
\newcommand{\by}{{\bf y}}

\begin{document}
\runauthor{PKU}
\begin{frontmatter}

\title{Lattice study on kaon nucleon scattering length%
      \ in the $I=1$ channel}

\author[PKU]{Guangwei Meng}
\author[PKU]{Chuan Miao}
\author[PKU]{Xining Du}
\author[PKU]{Chuan Liu}
\address[PKU]{School of Physics\\
              Peking University\\
              Beijing, 100871, P.~R.~China}

\begin{abstract}
Using the tadpole improved clover Wilson quark action on small,
coarse and anisotropic lattices, $KN$ scattering length in the
$I=1$ channel is calculated within quenched approximation. The
results are extrapolated towards the chiral and physical kaon mass
region. Finite volume and finite lattice spacing errors are also
analyzed and a result in the infinite volume and continuum limit
is obtained which is compatible with the experiment and the
results from Chiral Perturbation Theory.
\end{abstract}
\end{frontmatter}


\section{Introduction}

 Anisotropic lattices and improved actions have
 been extensively used in recent lattice QCD calculations.
 In our previous works, the following tadpole improved
 gluonic action on anisotropic lattices has been utilized:
 \ba
 \label{eq:gauge_action}
 S=&-&\beta\sum_{i>j} \left[
 {5\over 9}{TrP_{ij} \over \xi u^4_s}
 -{1\over 36}{TrR_{ij} \over \xi u^6_s}
 -{1\over 36}{TrR_{ji} \over \xi u^6_s} \right] \nonumber \\
 &-&\beta\sum_{i} \left[ {4\over 9}{\xi TrP_{0i} \over  u^2_s}
 -{1\over 36}{\xi TrR_{i0} \over u^4_s} \right] \;\;,
 \ea
 where $P_{0i}$ and $P_{ij}$ represents the usual temporal and
 spatial plaquette variable, respectively.
 $R_{ij}$ and $R_{i0}$ designates the $2\times 1$
 spatial and temporal Wilson loops, where, in order to eliminate
 the spurious states, we have restricted the coupling of
 fields in the temporal direction
 to be within one lattice spacing. The parameter $u_s$,
 which is taken to be the fourth root of the average
 spatial plaquette value, implements the tadpole improvement.
 With the tadpole improvement factor, the renormalization of the anisotropy
 (or aspect ratio) $\xi=a_s/a_t$ would be small.
 In this work, we will not differentiate the bare aspect
 ratio and the renormalized one in this work.
 Using this action, glueball and hadron spectra have been studied
 within quenched approximation \cite{colin97,colin99,%
 chuan01:gluea,chuan01:glueb,chuan01:canton1,chuan01:canton2,chuan01:india}.
 In our previous works, configurations generated from this
 improved action have also been utilized to calculate the
 $\pi\pi$ scattering lengths in the $I=2$ channel \cite{chuan02:pipiI2}.
 In this paper, we  extend our study to
 the kaon nucleon scattering lengths within quenched approximation.
 Unlike the case in pion pion scattering where the chiral expansion
 is quite reliable, chiral predictions in $KN$ scattering
 suffer from large
 corrections \cite{lee94:KN_chiral,savage94:KN_chiral,lee95:KN_chiral,kaiser01:KN_chiral}.
 On the experimental side, Martin used dispersion relation
 fits to determine the $KN$ scattering parameters \cite{martin81:KN}.
 Therefore, a lattice QCD calculation will offer an important
 and independent check on these results.

 Lattice calculations of hadron scattering lengths
 have been performed by various authors using conventional Wilson
 fermions on symmetric lattices without the
 improvement~\cite{gupta93:scat,fukugita95:scat,jlqcd99:scat}.
 In Ref.~\cite{fukugita95:scat}, a systematic study of the
 hadron scattering was performed, including pion-pion,
 kaon-nucleon, pion-nucleon and nucleon-nucleon scattering.
 However, this study was done only at some specific parameters
 and the results were not extrapolated towards the
 continuum and chiral limit.
 In later works, chiral and continuum limit was studied
 systematically in the case of pion-pion scattering,
 especially in the $I=2$
 channel \cite{jlqcd99:scat,chuan02:pipiI2,JLQCD02:pipi_length,CPPACS03:pipi_phase}
 It is known that, using the symmetric lattices and Wilson
 action, large lattices have to be simulated which require
 substantial amount of computing resources.
 In Ref.~\cite{chuan02:pipiI2}, we have shown that, pion-pion
 scattering lengths can be obtained using relatively small
 lattices with the help of tadpole improved Wilson quarks on
 anisotropic lattices. It is also possible to
 perform the chiral, infinite volume and
 continuum limits extrapolation.
 The final extrapolated result on the pion-pion scattering
 length in the $I=2$ channel is in good agreement with Chiral
 perturbation result, the experimental result and other
 lattice results using conventional Wilson
 fermions \cite{JLQCD02:pipi_length}.
 In this paper, similar techniques are applied and
 the kaon-nucleon scattering lengths in the $I=1$ channel
 is studied.

 The fermion action used in this study is the
 tadpole improved clover Wilson action on anisotropic
 lattices \cite{klassen99:aniso_wilson,chuan01:tune,chuan02:pipiI2}:
 \ba
 \label{eq:fmatrix}
 {\mathcal M}_{xy} &=&\delta_{xy}\sigma + {\mathcal A}_{xy}
 \nonumber \\
 {\mathcal A}_{xy} &=&\delta_{xy}\left[{1\over 2\kappa_{\max}}
 +\rho_t \sum^3_{i=1} \sigma_{0i} {\mathcal F}_{0i}
 +\rho_s (\sigma_{12}{\mathcal F}_{12} +\sigma_{23}{\mathcal F}_{23}
 +\sigma_{31}{\mathcal F}_{31})\right]
 \nonumber \\
 &-&\sum_{\mu} \eta_{\mu} \left[
 (1-\gamma_\mu) U_\mu(x) \delta_{x+\mu,y}
 +(1+\gamma_\mu) U^\dagger_\mu(x-\mu) \delta_{x-\mu,y}\right] \;\;,
 \ea
 where various coefficients in the fermion matrix
 ${\mathcal M}$ are given by:
 \ba
 \eta_i &=&\nu/(2u_s) \;\;, \;\;
 \eta_0=\xi/2 \;\;, \;\;\sigma=1/(2\kappa)-1/(1\kappa_{max})\;\;,
 \nonumber \\
 \rho_t &=& c_{SW}(1+\xi)/(4u^2_s) \;\;, \;\;
 \rho_s = c_{SW}/(2u^4_s) \;\;.
 \ea
 Among the parameters which appear in the fermion matrix, $c_{SW}$
 is the coefficient of the clover term and $\nu$ is the so-called
 bare velocity of light, which has to be tuned non-perturbatively
 using the single pion dispersion relations \cite{chuan01:tune}.
 Tuning the clover coefficients non-perturbatively is a difficult
 procedure. In this work, the tadpole improved tree-level
 value, namely $c_{SW}=1$, is used.

 In the fermion matrix~(\ref{eq:fmatrix}), the bare quark mass
 dependence is singled out into the parameter $\sigma$ and the
 matrix ${\mathcal A}$ remains unchanged if the bare quark mass is
 varied.  This shifted structure of the matrix ${\mathcal M}$ can be utilized
 to solve for quark propagators at various values of valance quark mass
 $m_0$ (or equivalently $\kappa$)
 at the cost of solving only the lightest valance
 quark mass value at $\kappa=\kappa_{\max}$,
 using the so-called  Multi-mass Minimal
 Residual ($M^3R$ for short) algorithm %
 \cite{frommer95:multimass,glaessner96:multimass,beat96:multimass}.
 This is particularly advantageous in a quenched calculation
 since one needs the results at various quark mass values
 to perform the chiral extrapolation.

 This paper is organized as follows.
 In Section~\ref{sec:theory}, the method to calculate $KN$ scattering lengths is
 briefly reviewed. The $KN$ correlation functions used in our calculation
 are defined. In Section~\ref{sec:simulation},
 some simulation details are described.
 In Section~\ref{sec:extrap}, our results
 of the $KN$ scattering lengths obtained on lattices
 of various sizes and lattice spacings are extrapolated
 towards the chiral limit. Physical strange quark mass
 is also determined from the $K$-input.
 Finite size effects are then studied and the
 infinite volume limit are taken. Finally, our lattice
 results are extrapolated towards the continuum limit.
 Comparisons with the experimental value and
 values from chiral perturbation
 theory at various orders are also discussed.
 In Section~\ref{sec:conclude}, we conclude with some general remarks.

 \section{Formulation to extract the scattering length}
 \label{sec:theory}

 In order to calculate the elastic scattering lengths for hadron-hadron
 scattering on the lattice,
 or the scattering phase shifts in general, one uses L\"uscher's
 formula which relates the exact energy level of two hadron states
 in a finite box to the elastic scattering phase shift in the continuum.
 In the case of $KN$ scattering at
 zero relative three momentum, this formula amounts to a relation
 between the {\em exact} energy $E^{(I)}_{KN}$ of the
 $KN$ system in a finite
 box of size $L$ with isospin $I$, and
 the corresponding scattering length
 $a^{(I)}_0$ in the continuum.
 This formula reads \cite{luscher86:finiteb}:
 \be
 \label{eq:luescher}
 E^{(I)}_{KN}-(m_K+m_N)=-\frac{2\pi a^{(I)}_0}{\mu_{KN} L^3}
 \left[1+c_1\frac{a^{(I)}_0}{L}+c_2(\frac{a^{(I)}_0}{L})^2
 \right]+O(L^{-6}) \;\;,
 \ee
 where $c_1=-2.837297$, $c_2=6.375183$ are numerical constants
 and $\mu_{KN}=m_Km_N/(m_K+m_N)$ is the reduced mass of the
 $KN$ system. In this paper, the $KN$ scattering length $a^{(1)}_0$
 in the $I=1$ channel will be studied.

 To measure the hadron mass values $m_K$, $m_N$ and
 to extract the energy shift $\delta E^{(1)}_{KN}$, we construct the correlation
 functions from the corresponding operators in the $I=1$ channel.
 In this work, the hadron states that need to be measured include pion,
 kaon, rho, nucleon and $KN$ states.
 The local operators which create a single pion with
 appropriate isospin values are:
 \ba
 \pi^+(\bx,t)&=&-\bar{d}(\bx,t)\gamma_5u(\bx,t) \;\;,\;\;
 \pi^-(\bx,t)=\bar{u}(\bx,t)\gamma_5d(\bx,t) \;\;,\;\;
 \nonumber \\
 \pi^0(\bx,t)&=&{1 \over \sqrt{2}}
 [\bar{u}(\bx,t)\gamma_5u(\bx,t)-\bar{d}(\bx,t)\gamma_5d(\bx,t)]\;\;,
 \ea
 where $u(\bx,t)$, $d(\bx,t)$, $\bar{u}(\bx,t)$ and
 $\bar{d}(\bx,t)$ are basic local quark fields corresponding to $u$ and
 $d$ quarks, respectively.
 The operators which correspond to zero momentum single pions are:
 \be
 \pi^a_0(t)={1 \over L^{3/2}} \sum_{\bx} \pi^a(\bx,t) \;\;,
 \ee
 where the flavor of pions $a=+,-,0$ and $L^3$ is the three volume
 of the lattice. Zero momentum one pion correlation function
 can then be formed:
 \be
 C_\pi(t)=<\pi^{\bar{a}}_0(t)\pi^a_0(0)> \;\;.
 \ee
 Similarly, one can construct the kaon correlation functions
 from kaon operators. In the $I=1$ channel, only $K^+$ and
 proton operators are needed:
 \ba
 K^+(\bx,t)&=&-\bar{s}(\bx,t)\gamma_5u(\bx,t) \;\;,
 \nonumber \\
 p(\bx,t)&=& \epsilon_{abc}
 \left[u^T_a(\bx,t)(C^{-1}\gamma_5)d_b(\bx,t)\right]
 u_c(\bx,t)\;.
 \ea
 The proton operator given above is itself a Dirac spinor.
 The zero momentum proton-proton correlation function
 is obtained by projecting out the positive parity
 part of the zero momentum spinor operator and taking
 the trace in Dirac space:
 \be
 C_N(t)=\sum_{\bx,\by}\left\langle p(\bx,0)
 \left({1+\gamma_0\over 2}\right)
 p(\by,t)\right\rangle \;.
 \ee
 From the large $t$ behavior of the correlation functions,
 the pion mass $m_\pi$, the kaon mass $m_K$ and
 the nucleon mass $m_N$ are obtained.

 Two hadron operators are used in the
 construction of two hadron correlation functions. For example,
 the $KN$ operator in the $I=1$ channel is given by:
 \be
 O^{I=1}_{KN}(t) = K^+_0(t)p_0(t+1)\;\;,
 \bar{O}^{I=1}_{KN}(t) = K^-_0(t)\bar{p}_0(t+1)\;\;,
 \ee
 where $K^+_0(t)$ and $p_0(t)$ are the zero momentum kaon and
 proton operators respectively.
 From these operators, two hadron correlation functions
 can be constructed. The $KN$ correlation
 function in the $I=1$ channel reads:
 \be
 C^{I=1}_{KN}(t)=<\bar{O}^{I=1}_{KN}(0)
 \left({1+\gamma_0\over 2}\right)O^{I=1}_{KN}(t)> \;\;.
 \ee
 Numerically, it is more advantageous to construct
 the ratio of the correlation functions defined above:
 \be
 {\mathcal R}^{I=1}(t) = C^{I=1}_{KN}(t) / (C_K(t)C_N(t)) \;\;.
 \ee
 This ratio thus exhibits the following asymptotic behavior for
 large $t$:
 \be
 {\mathcal R}^{I=1}(t) \stackrel{t >>1}{\sim}
 e^{-\delta E^{(1)}_{KN}t} \;\;,
 \ee
 with $\delta E^{(1)}_{KN}=E^{(1)}_{KN}-m_K-m_N$
 is the energy shift in this channel which directly enters L\"uscher's
 formula~(\ref{eq:luescher}).
 This formula is utilized for
 small $\delta E^{(1)}_{KNi} t$ values (large enough $L$)
 in the calculation where the signal to noise ratio is good.
 In this case, one could equally use the linear fitting function:
 \be
 \label{eq:linear_fit}
 {\mathcal R}^{I=1}(t) \stackrel{T >> t >>1}{\sim} 1-\delta E^{(1)}_{KN}t
 \;\;,
 \ee
 to determine the energy shift $\delta E^{(1)}_{KN}$.
 Similar procedure has also been used in lattice studies
 of pion-pion scattering lengths by various groups.

 $KN$ correlation function, or equivalently, the ratio
 ${\mathcal R}^{I=1}(t)$ constructed above can be
 transformed into products of quark propagators using
 Wick's theorem.
 \begin{figure}[thb]
 \begin{center}
 \includegraphics[width=12.0cm,angle=0]{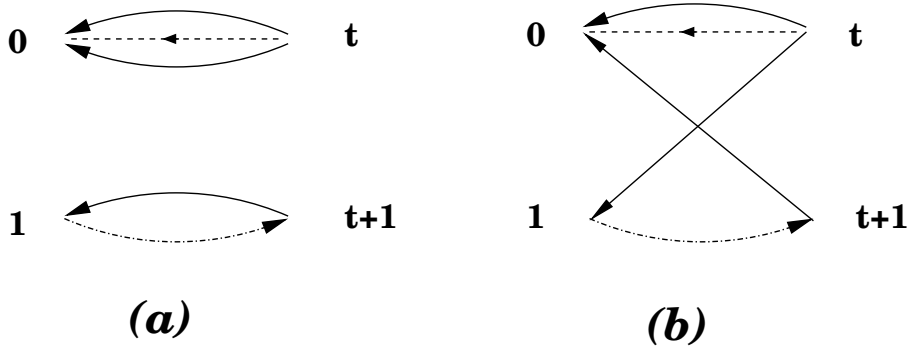}
 \end{center}
 \caption{The diagrams that contribute to $KN$ scattering in
 the $I=1$ channel. The solid lines represent $u$-quark
 propagators while dashed lines are $d$-quark propagators.
 The dashed-dotted lines designates the strange quark
 propagator. Note  that diagram (b) in fact contains
 two different contractions.
 \label{fig:diagram}}
 \end{figure}
 These contributions are represented by diagrams
 as shown in Fig.~\ref{fig:diagram}, where the solid, the dashed
 and the dashed-dotted lines represent the $u$, $d$
 and $s$ quark propagators,
 respectively.\footnote{In this work, the $u$ and $d$ quark
 are assumed to be degenerate in mass.}%
 The $KN$ correlation function in the
 $I=0$ channel is, however, more complicated which
 involves vacuum diagrams that require to compute
 the quark propagators for wall sources placed at
 {\em every} time-slice, a procedure which is more
 time-consuming than in the $I=1$ channel.

 \section{Simulation details}
 \label{sec:simulation}

 Simulations are performed on a PC based cluster with $30$ nodes.
 Configurations are generated using the pure
 gauge action~(\ref{eq:gauge_action})
 for $6^340$, $8^340$ and $10^350$ lattices with the gauge
 coupling $\beta=1.7$, $1.9$, $2.2$ and $2.4$. The spatial lattice
 spacing $a_s$ is roughly between $0.2$fm and $0.4$fm while
 the physical size of the lattice ranges from
 $1.2$fm to $4.0$fm. For each set of parameters, several hundred
 decorrelated gauge field configurations are used to measure
 the fermionic quantities. Statistical errors are all analyzed
 using the usual jack-knife method. Basic information of these
 configurations is summarized in Table~\ref{tab:parameters}.
 \begin{table}[htb]
 \caption{Simulation parameters for lattices studied in
 this work. Input aspect ratio parameter $\xi$ is fixed to be $5$
 for all lattices being studied.
 The approximate spatial lattice
 spacing $a_s$ in physical units as obtained from
 \cite{colin99,chuan01:india} is also indicated.
 Also listed are the maximum value of the hopping parameter
 $\kappa_{\max}$. In the last three columns, we listed
 the physical up/down quark hopping parameters, the physical
 strange quark hopping parameters and the
 results for the factor $F=a^{(1)}_0m^2_\rho/\mu_{KN}$ at physical
 quark mass values, respectively. In the rows labelled as
 "Extrapolated", we have also listed the values of $F$ after the
 infinite volume extrapolation.
 \label{tab:parameters}}
 \begin{center}
 \begin{tabular}{|c|c|c||c|c|c|}
 \hline
  Lattice & No. confs  & $\kappa_{\max}$ %
 & $\kappa^{(phy)}_u$ & $\kappa^{(phy)}_s$ & $F=a^{(1)}_0m^2_\rho/\mu_{KN}$ \\
 \hline
 \multicolumn{6}{|c|}
 {$\beta=2.4$, $u^4_s=0.4090$, $a_s\simeq 0.22$fm, $\nu$=0.93}
 \\ \hline
 $ 6^340$ &  $288$ & $0.0610$ & $0.06179(2)$  & $0.06120(6)$ & $-0.7(15)$\\
 $ 8^340$ &  $192$ & $0.0610$ & $0.061791(9)$ & $0.06113(4)$ & $-1.6(6)$\\
 $10^350$ &  $204$ & $0.0605$ & $0.061636(9)$ & $0.06113(5)$ & $-3.3(9)$\\
 \hline
 Infinite &   &   &   &   & $-3.3(11)$\\
 \hline
 \multicolumn{6}{|c|}
 {$\beta=2.2$, $u^4_s=0.378$, $a_s\simeq 0.27$fm, $\nu$=0.95}
 \\ \hline
 $ 6^340$   &  $256$ & $0.0605$ & $0.06116(1)$ & $0.06020(2)$ & $-1.9(3)$\\
 $ 8^340^*$ &  $224$ & $0.0600$ & $0.06113(1)$ & $0.06041(5)$ & $-3.1(9)$\\
 $10^350$ &  $184$ & $0.0605$ & $0.061230(7)$ & $0.06024(2)$ & $-3.3(13)$\\
 \hline
 Infinite &   &   &   &   & $-3.9(11)$\\
 \hline
 \multicolumn{6}{|c|}
 {$\beta=1.9$, $u^4_s=0.3281$, $a_s\simeq 0.35$fm, $\nu$=0.93}
 \\ \hline
 $ 6^340^*$ &  $304$ & $0.0600$ & $0.06087(1)$  & $0.05945(2)$  & $-2.3(7)$\\
 $ 8^340^*$ &  $160$ & $0.0595$ & $0.06073(1)$  & $0.05969(5)$  & $-2.8(46)$ \\
 $10^350^*$ &  $252$ & $0.0595$ & $0.060722(6)$ & $0.05964(2)$ & $-4.0(44)$ \\
 \hline
 Infinite &   &   &   &   & $-4.1(46)$\\
 \hline
 \multicolumn{6}{|c|}
 {$\beta=1.7$, $u^4_s=0.295$, $a_s\simeq 0.39$fm, $\nu$=0.90}
 \\ \hline
 $ 6^340^*$ &  $288$ & $0.0590$ & $0.06027(1)$ & $0.05895(3)$ & $-2.9(25)$\\
 $ 8^340^*$ &  $240$ & $0.0585$ & $0.06019(1)$ & $0.05906(4)$ & $-3.1(58)$\\
 $10^350^*$ &  $250$ & $0.0580$ & $0.06007(2)$ & $0.05914(9)$ & $-3.4(62)$\\
 \hline
 Infinite &   &   &   &   & $-4.6(64)$\\
 \hline
 \end{tabular}
 \end{center}
 \end{table}

 Quark propagators are measured using the Multi-mass Minimal
 Residue algorithm for several different values of bare quark mass.
 Periodic boundary condition is applied to all three
 spatial directions while in the temporal direction, Dirichlet
 boundary condition is utilized. In this calculation, it is
 advantageous to use the wall sources which enhance
 the signal \cite{gupta93:scat,fukugita95:scat,jlqcd99:scat,chuan02:pipiI2}.
 Values of the maximum hopping parameter $\kappa_{\max}$,
 which corresponds to the lowest valance quark mass, are also
 listed in Table~\ref{tab:parameters}. Typically, a few hundred
 Minimal Residual iterations are needed to obtain the solution
 vector for a given source. On small lattices, in
 particular those with a low value of $\beta$, the hopping
 parameter has to be kept relatively far away from its
 critical value in order to avoid the appearance of
 exceptional configurations.
 The parameter $\nu$, also
 known as the bare velocity of light, that enters the fermion
 matrix~(\ref{eq:fmatrix}) is determined non-perturbatively
 using the single pion dispersion relations as described in
 Ref.~\cite{chuan01:tune} and Ref.~\cite{chuan02:pipiI2}.
 The optimal value of $\nu$ which corresponds to each
 parameter set of the simulation is also tabulated in
 Table~\ref{tab:parameters}

 Single pion, kaon, rho and nucleon mass values
 are obtained from the plateau of their corresponding effective
 mass plots. The fitting interval is automatically
 chosen by minimal $\chi^2$ per degree of freedom.
 Due to the usage of finer lattice spacing in
 the temporal direction, good plateau behavior was observed
 in these effective mass plots.
 Therefore, contaminations from
 excited states should be negligible.
 These mass values will be utilized in the chiral
 extrapolation.

 \begin{figure}[thb]
 \begin{center}
 \includegraphics[height=12.0cm,angle=0]{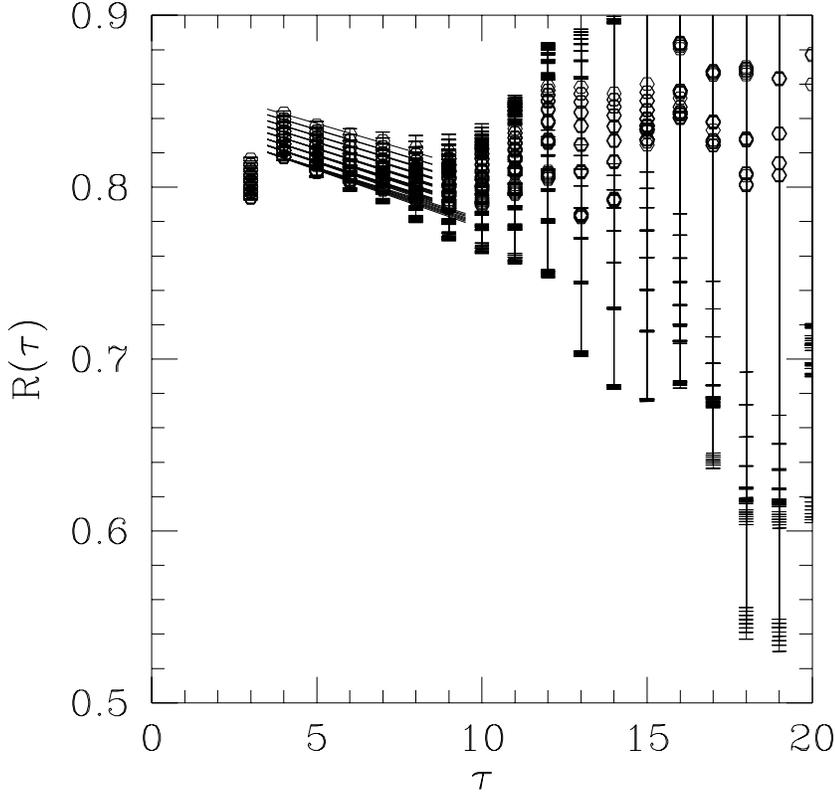}
 \end{center}
 \caption{The ratio $R^{(1)}(t)$ as a function of $t$ obtained from the
 $KN$ correlation functions for $8$ different values of the bare quark mass.
 The lattice size is $8^340$ and the gauge coupling
 $\beta=2.4$. The optimal bare velocity of light is taken to be $\nu=0.93$.
 The straight lines are linear fits according to
 Eq.~(\ref{eq:linear_fit}), from which the energy
 shifts $\delta E\equiv E^{(1)}_{KN}-m_K-m_N$ are extracted
 for all possible values of $(\kappa_u, \kappa_s)$.}
 \label{fig:KN_ratio}
 \end{figure}
 $KN$ correlation functions and the ratio
 ${\mathcal R}^{(1)}(t)$ are
 constructed from products of suitable quark propagators
 according to Wick's theorem. Both the Direct and the Cross
 contributions are included. For the ratio ${\mathcal R}^{(1)}(t)$,
 we obtain clear signal for all our data sets.
 Since a kaon consists of one up (down) and one strange
 quark whose mass is different from the up and down quarks,
 all physical quantities involving the kaon will depend on
 two quark masses. In terms of the hopping parameters,
 the $KN$ correlation functions depends on a pair
 of hopping parameters: $(\kappa_u,\kappa_s)$.
 In Fig.~\ref{fig:KN_ratio} we show the ratio
 ${\mathcal R}^{(1)}(t)$ for different pairs of $\kappa$ values
 as a function of the temporal separation $t$,
 together with the corresponding linear fit~(\ref{eq:linear_fit}) for
 $\beta=2.4$ on $8^340$ lattices.
 The starting and ending positions of the straight lines
 indicate the appropriate fitting
 range which minimizes the $\chi^2$ per degree of freedom.

 After obtaining the energy shifts $\delta E^{(1)}_{KN}$,
 these values are substituted into L\"uscher's formula to solve
 for the scattering length $a^{(1)}_0$ for all possible
 hopping parameter pairs: $(\kappa_u,\kappa_s)$, which
 corresponds to the up (down) and the strange
 bare quark mass values respectively.
 This is done for lattices of all sizes being
 simulated and for all values of $\beta$.
 From these results, attempts are made to perform
 an extrapolation towards the chiral, infinite volume
 and continuum limit.

 \section{Extrapolation towards chiral, infinite volume and continuum limits}
 \label{sec:extrap}

 The chiral extrapolations of physical quantities involving both
 the up (down) and the strange quarks consists of two
 steps. In the first step, the bare strange quark mass, or equivalently
 the corresponding hopping parameter $\kappa_s$, is kept
 fixed while the hopping parameter of the up/down quark, $\kappa_u$,
 is brought to their physical value $\kappa^{(phy)}_u$.
 The precise value of $\kappa^{(phy)}_u$ can be obtained by
 inspecting the chiral behavior of the pseudo-scalar (pion)
 and the vector meson (rho) mass values.
 In the second step, one fixes the up/down quark
 hopping parameter at its physical value obtained in the first step
 and extrapolate/interpolate in
 the strange quark mass. The physical strange quark hopping parameter
 $\kappa^{(phy)}_s$ is determined by demanding the mass of
 the kaon being exactly at its physical value ($493$MeV).
 In practice, this could be accomplished by, for example,
 inspecting the kaon to rho mass ratio.
 This is also known as the $K$-input.
 After these two steps, we would be able to obtain physical
 quantities at the physical up/down and strange quark mass values.
 In Table~\ref{tab:parameters}, we have listed the physical hopping
 parameters $\kappa^{(phy)}_u$  and
 $\kappa^{(phy)}_s$ for the up/down quarks and the strange quark.
 Note that one only needs the pion, kaon and rho correlation
 functions to determine the physical hopping parameters of the quarks.

 We now come to the chiral extrapolation of the scattering length.
 In the chiral limit, the $KN$ scattering length in
 the $I=1$ channel is given by the current algebra result:
 \be
 \label{eq:CA}
 a^{(1)}_0 =-{1 \over 4 \pi} {\mu_{KN} \over f^2_K} \;\;,
 \ee
 where $a^{(1)}_0$ is the $KN$ scattering length in
 the $I=1$ channel and $f_K\sim 113$MeV is the kaon decay constant.
 Chiral Perturbation Theory to one-loop order gives an
 expression which includes the next-to-leading order
 contributions \cite{kaiser01:KN_chiral}.
 Unlike the chiral perturbation theory results for the mesons,
 the chiral expansion for baryons also receives corrections
 which are suppressed by the meson to baryon mass ratios.
 Since the mass of the kaon is not light,
 one typically obtains large corrections from the next chiral order compared
 with the previous one. In the case of the $KN$ scattering length
 $a^{(1)}_0$, the next order correction is $114$\% with the
 opposite sign and the next-to-next order is about $94$\% with
 the same sign of the tree-level. This is a signal that
 one really has no control over the expansion.

 To perform the chiral extrapolation of the scattering length,
 it is more convenient to use the quantity
 $F=a^{(1)}_0m^2_\rho/\mu_{KN}$, which in the chiral limit reads:
 \be
 \label{eq:chiral}
 F\equiv {a^{(1)}_0m^2_\rho \over \mu_{KN}}
 =-{1 \over 4 \pi} {m^2_\rho \over f^2_K}
 \sim  -3.686\;\;,
 \ee
 where the final numerical value is obtained
 by substituting in the experimental values for
 $m_\rho\sim 770$MeV and $f_K\sim 113$MeV.
 The factor $F$ can be calculated on the lattice with good
 precision {\em without} the lattice calculation of meson
 decay constants. The error
 of the factor $F$ obtained on the lattice will mainly
 come from the error of the scattering length $a^{(1)}_0$, or
 equivalently, the energy shift
 $\delta E^{(1)}_{KN}$ and nowhere else.
 Since we have calculated the factor $F$ for several different
 values of valance quark mass, we could make a chiral
 extrapolation and extract the corresponding results in the chiral limit.

\begin{figure}[thb]
 \begin{center}
 \includegraphics[height=12.0cm,angle=0]{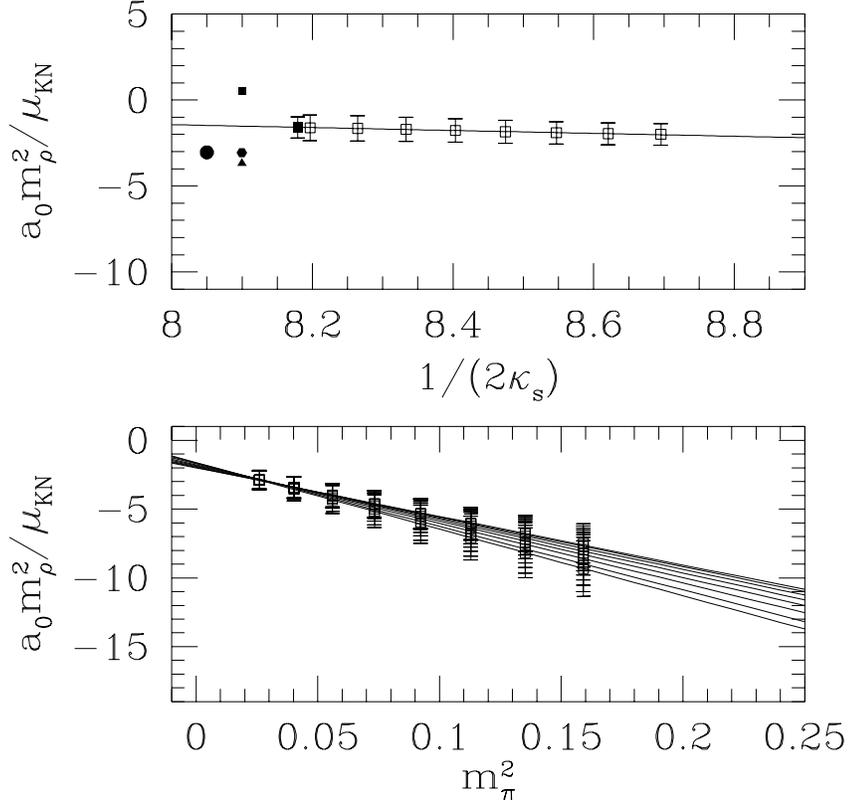}
 \end{center}
 \caption{Chiral extrapolation for the quantity
 $F\equiv a^{(1)}_0m^2_\rho/\mu_{KN}$ for our simulation results
 at $\beta=2.4$ on $8^340$ lattices. In the lower panel, the
 factor $F$ as defined in Eq.~(\ref{eq:chiral}) are plotted
 versus the pion mass squared, for fixed
 strange quark mass. The straight lines represent
 the corresponding linear fit in $m^2_\pi$
 for the data. In the upper panel of the plot, the extrapolated quantity $F$
 is plotted (open squares)
 as a function of strange quark mass parameter $1/(2\kappa_s)$.
 The straight line represents the linear
 interpolation/extrapolation to the physical strange quark mass, where
 the result is depicted as a solid square.
 As a comparison, the
 corresponding semi-experimental result \cite{martin81:KN}
 for this quantity is shown as a filled circle. Chiral perturbation theory
 results are also shown as a filled triangle (tree), a filled square (next) and
 a filled hexagon (next-to-next), respectively.
 \label{fig:chiral_extrapolation}}
 \end{figure}

 First, we extrapolate the factor $F$ in the up/down quark mass
 values. In this step, we have adopted a simple linear extrapolation.
 the fitting range of the extrapolation is self-adjusted by the
 program to yield a minimal $\chi^2$ per degree of freedom.
 Second, the extrapolated values for the factor $F$ are then
 extrapolate/interpolate in the strange quark mass. Our hopping
 parameters are chosen such that the physical strange quark
 mass is either within the range of our choice or very close
 to avoid a long extrapolation in the strange quark mass.
 In Fig.~\ref{fig:chiral_extrapolation}, we have shown the
 two step chiral extrapolation for one of our simulation points.
 In the lower panel of Fig.~\ref{fig:chiral_extrapolation},
 we have plotted the results for the factor $F$ for all possible
 pairs of quark mass values. They are arranged according to
 different up/down quark hopping parameter. The straight lines
 show the extrapolation in $m^2_\pi$ (the first
 step extrapolation). The physical up/down quark hopping parameter
 are read from the pion and rho correlation function fits which
 are tabulated in Tab.~\ref{tab:parameters}.
 The resulting factor $F$ after the first extrapolation
 are plotted in the upper panel of
 Fig.~\ref{fig:chiral_extrapolation} versus
 the strange quark mass parameter $1/(2\kappa_s)$.
 The straight line is the linear  extrapolation/interpolation
 (second extrapolation) and the final result
 is also depicted as a solid square at the physical strange
 quark mass. Current algebra result and the result from chiral perturbation theory
 are also shown as a filled triangle (tree level),
 a filled square (next order) and a filled hexagon (next-to-next),
 respectively. The result from Ref.~\cite{martin81:KN} is
 shown as a filled circle. To avoid crowdedness, these symbols are shifted
 to the left. The fitting quality for the factor $F$ are reasonable.
 The quality for the chiral extrapolation of other simulation points
 are similar. As is seen, the linear fit gives a
 reasonable modelling of the data. The divergent contributions
 and other higher order corrections from quenched chiral perturbation
 theory seem to be numerically small for the lattices being simulated.

 After the chiral extrapolation, we now turn to study
 the finite volume effects of the simulation. According
 to formula~(\ref{eq:luescher}), the quantity $F$ obtained
 from finite lattices differs from
 its infinite volume value by corrections of the
 form $1/L^3$. However, the situation might be
 different in a quenched calculation, which
 happened in the case of pion-pion
 scattering \cite{fukugita95:scat,bernard96:quenched_scat}.
 In the case of $KN$ scattering length,
 a similar calculation using quenched chiral perturbation theory
 is still not available.
 Therefore, we will be using a linear extrapolation in $1/L^3$,
 as suggested by L\"uscher's formula.
 In Table.~\ref{tab:parameters},
 we have listed the infinite volume extrapolation for
 our  simulation points
 at $\beta=2.4$, $2.2$, $1.9$ and $1.7$.

 \begin{figure}[thb]
 \begin{center}
 \includegraphics[height=12.0cm,angle=0]{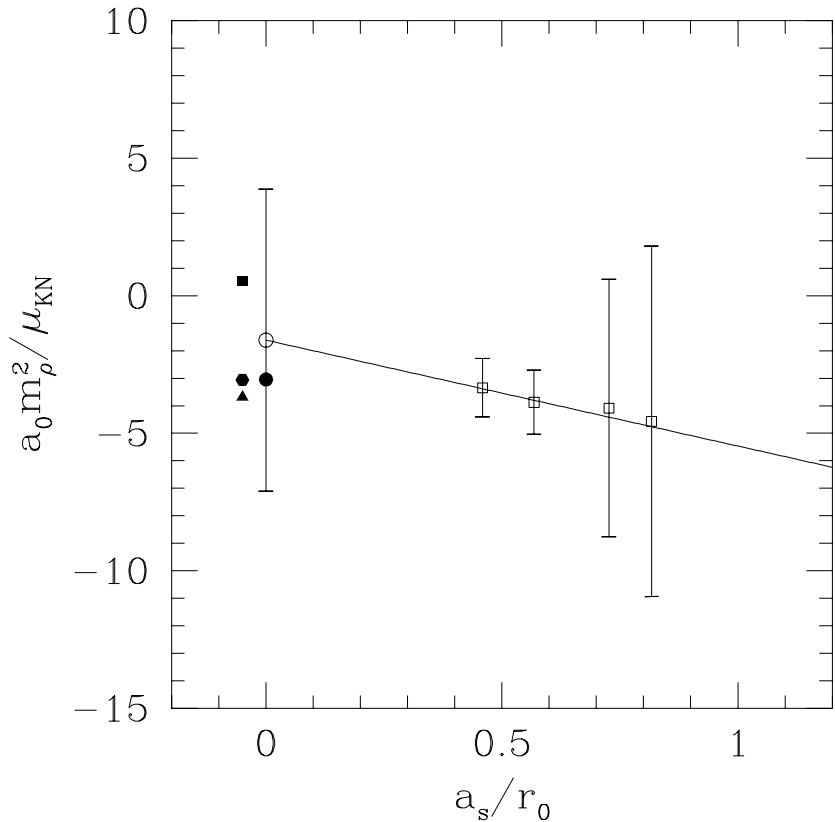}
 \end{center}
 \caption{Continuum extrapolation for the quantity
 $a^{(1)}_0m^2_\rho/\mu_{KN}$ obtained from our simulation results
 at $\beta=2.4$, $2.2$, $1.9$ and $1.7$.
 The straight line represents the linear extrapolation in $a_s/r_0$.
 The extrapolated results are also shown, together with the
 experimental result from Ref.~\cite{martin81:KN} indicated
 by the filled circle. For comparison, the results from Chiral perturbation
 theory are also shown as the points at $a_s=0$ respectively.
 \label{fig:continuum_extrapolation}}
 \end{figure}
 Finally, we can make an extrapolation towards the
 continuum limit by eliminating the finite lattice
 spacing errors. Since we have used the tadpole
 improved clover Wilson action, all physical quantities
 differ from their continuum counterparts by
 terms that are proportional to $a_s$. The physical
 value of $a_s$ for each value of $\beta$ can be
 found from Ref. \cite{colin99,chuan01:india}, which
 is also included in Table~\ref{tab:parameters}.
 This extrapolation is shown in
 Fig.~\ref{fig:continuum_extrapolation} where
 the results from the chiral and infinite volume
 extrapolation discussed above are indicated as
 data points in the plot for all $4$ values of
 $\beta$ that have been simulated.
 The straight line shows
 the extrapolation towards the $a_s=0$ limit and the
 extrapolated results are also shown as solid squares together with
 the experimental result from Ref.\cite{martin81:KN}
 which is shown as the filled circle. For comparison,
 the results from chiral perturbation
 theory ar various orders are also shown as the filled triangle, the filled square and
 the filled hexagon at $a_s=0$, respectively.
 It is seen from the figure that our extrapolated result is
 negative, with a large error bar. Therefore, we can only say
 that, from our preliminary study we obtain a
 result which is compatible with the experimental result and results
 from chiral perturbation theory. The error of our calculation
 can be reduced by further studies with more statistics and
 on lattices with smaller lattice spacing in the future.
 To summarize, we obtain from the linear extrapolation the following
 result for the quantity
 $F=a^{(1)}_0m^2_\rho/\mu_{KN}=(-1.6\pm 5.4)$.
 If we substitute in the physical values, we obtain the
 $KN$ scattering length in the $I=1$ channel:
 $a^{(1)}_0= (-0.17 \pm 0.59) \mbox{fm}$, which is to be compared
 with the experimental result of $(-0.33) \mbox{fm}$ and the
 chiral result of $(-0.398+0.457-0.389)\mbox{fm}$ at the first,
 second and third chiral order, respectively.

 \section{Conclusions}
 \label{sec:conclude}

 In this paper, we have calculated kaon nucleon scattering
 lengths in isospin $I=1$ channel using quenched
 lattice QCD. It is shown that such a calculation
 is feasible using relatively small, coarse and anisotropic lattices.
 The calculation is done using the
 tadpole improved clover Wilson action on anisotropic
 lattices. Simulations are performed on lattices
 with various sizes, ranging from $1.2$fm to about
 $4$fm and with four different values of lattice spacing.
 Quark propagators are measured with different
 valence quark mass values. These enable us to
 explore the finite volume and the finite
 lattice spacing errors in a systematic fashion.
 The infinite volume extrapolation is made.
 The lattice result for the scattering length is
 extrapolated towards the chiral
 and continuum limit where a result consistent with
 the experiment and the Chiral Perturbation Theory is found.
 We believe that, using the method described in this exploratory
 study, more reliable results on $KN$
 scattering lengths could be obtained with more statistics
 and with simulations on finer lattices.

 \section*{Acknowledgments}

 This work is supported by the National Natural
 Science Foundation (NFS) of China under grant
 No. 90103006, No. 10235040 and supported by the Trans-century fund from Chinese
 Ministry of Education. C. Liu would like
 to thank Prof.~H.~Q.~Zheng and Prof.~S.~L.~Zhu for helpful discussions.


\end{document}